\documentclass[fp,twocolumn]{jpsj3}
\usepackage{txfonts}
\usepackage{graphicx}
\usepackage[usenames]{color}

\title{Higher-order tensor renormalization group approach to lattice glass model}

\author{
Kota Yoshiyama$^1$\thanks{yoshiyama@huku.c.u-tokyo.ac.jp},
Koji Hukushima$^{1,2}$
}
\inst{$^1$Graduate School of Arts and Sciences, The University of Tokyo, Komaba
3-8-1,  Meguro, Tokyo 153-8902, Japan \\
$^2$Komaba Institute for Science, The University of Tokyo, Komaba 3-8-1, Meguro, Tokyo 153-8902, Japan
} 

\abst{
In this study, the higher-order tensor renormalization group (HOTRG) method is applied to a lattice glass
model that has local constraints on the occupation number of neighboring
particles represented by many-body interactions. This model exhibits first- and second-order transitions depending on a certain model parameter.
The results obtained by using the HOTRG method for the model were confirmed to be consistent with those obtained by a Markov-chain Monte Carlo (MCMC) method for systems of relatively small sizes.
The transition points are accurately estimated by the HOTRG calculation for the systems of large sizes, which is challenging to
perform using the MCMC method.
These results demonstrate that the HOTRG method can be an efficient method for studying systems with many-body interactions.
}

\usepackage{bm}
\begin{document}
\maketitle

\section{Introduction}
Tensor network (TN) representations provide a powerful tool for studying quantum many-body systems.
Beginning with the density matrix renormalization group
method\cite{White1992} for one-dimensional quantum systems, TN methods
have been applied to higher-dimensional
systems\cite{Verstraete2004,Xie2014}.
In particular, such methods are expected to be effective
for quantum frustrated models without suffering from the negative sign problem encountered in quantum Monte Carlo methods.

The tensor renormalization group (TRG) method was proposed \cite{Levin2007} for classical statistical mechanical models,
and is based on the real space renormalization group method using the TN
\cite{Efrati2014a}. This method has recently attracted considerable research interest
as an efficient and non-perturbative numerical method for calculating
the partition function
in classical finite-dimensional lattice systems.
While Markov-chain Monte Carlo (MCMC) methods have been successfully
applied to a large number of quantum and classical statistical mechanical models,
the slowing down associated with phase transitions and extremely slow
relaxation in glassy systems have significantly hindered the application of the MCMC
methods to interesting systems.
However, in contrast to the MCMC methods, the TRG method is considered
to have essentially no such difficulty because it is not based on
importance sampling.

In recent years, intensive researches on TRG methods have been conducted
from various perspectives, such as their application to models in
two or higher dimensions\cite{Xie2012}, improving numerical
accuracy\cite{Gu2009, Evenbly2015}, and reducing the amount of computation required\cite{Adachi2019}.
As a result,
the TRG methods now enable calculations that are challenging to perform using
MCMC methods.
On the contrary, TRG methods are used only for a limited number of
models represented by two-body interactions, such as the Ising model,
and have not been applied to many models.
One of such unstudied but interesting models is a lattice gas model
proposed by Biroli and M\'{e}zard\cite{Biroli2002}, called the lattice glass model.
The model has local constraints on occupation numbers represented by
$k$-body interactions with $k\geq 3$.
Systems with many-body interactions cannot be expressed by
the conventional TN construction methods, except for some simple
systems, and TRG methods have not been applied to them.
Therefore, in this paper, we construct a TN for the Biroli-M\'{e}zard
(BM) model \cite{Biroli2002} and study the phase transition of the model
using TRG based on the TN.

The remainder of this paper is organized as follows:
In Sec.~\ref{sec:model}, we explain
the BM model, which is a lattice glass model with
many-body interactions. Sec.~\ref{sec:method} introduces the TN representation for the model and explains the higher-order tensor renormalization group
(HOTRG), a kind of the TRG methods we use in this work. Sec.~\ref{sec:results} presents our numerical results of the
model using the HOTRG, and
Sec.~\ref{sec:discussions} is devoted to discussions of these results and a summary of this paper.

\section{Model}
\label{sec:model}
The model discussed here is a lattice gas model with many-body interactions.
The particle occupation variable $n_i$ on each site of a given lattice is
defined as $n_i\in \{0, 1\}$, depending on the site $i$ that is being
(un)occupied. In the BM model, any particle configuration on a lattice is specified by a
vector of the occupation variable $\bm{n}=(n_1,n_2,\dots,n_N)$, where $N$
is the total number of sites.
The occupation is restricted by a hard constraint that any occupied site
with $n_i=1$ can have at most $l$ neighboring occupied sites.
This type constraint often yields two- or many-body interactions
between particles depending on the model parameter $l$.
The BM model defined on a random graph has been extensively studied
using the replica method and the cavity method. The results have shown that the
model exhibits a thermodynamic glass transition in the mean-field
limit\cite{Jorg2008, Rivoire2004}.
Unlike the mean-field model on a random graph, the finite-dimensional BM
model generally has a strong tendency to crystallize at the high
density. While the BM model of the honeycomb lattice is still under
discussion\cite{Hukushima2010}, it seems that the model on a square lattice does not have a glass transition.

In this paper, we study the BM model defined on the square lattice with a linear dimension $L$ in two dimensions (2D), in particular, focusing on the cases $l=0$ and $l=2$.
For each site $i$, the set of its four nearest neighboring sites is denoted by
$N(i)=\{i_u,i_l,i_d,i_r\}$, where the subscripts $u,l,d,r$ mean up,
left, down and right, respectively. The constraint for $n_i$ is expressed using the
occupation variables $n_{i_k}$ with $i_k\in N(i)$ as
\begin{equation}
 \forall i ~~~\sum_{k\in\{u,l,d,r\}} n_i n_{i_k}\leq l.
  \label{eqn:BMconstraint}
\end{equation}

The properties of equilibrium states of the system for a given chemical potential
$\mu$ and inverse temperature $\beta$ are described by the
ground-canonical partition function:
\begin{equation}
 Z(\mu,L) = \sum_{\bm{n}}\prod_i^N\left(\exp\left(\beta\mu
					 n_i\right)C\left(l-\sum_k
					 n_in_{i_k}\right)\right)
 \label{eqn:BMmodel}
\end{equation}
where the sum is taken over all the possible configurations $\bm{n}$, and
$C(x)$ is the Heaviside step function.
In the following, the unit of chemical potential is set as $1/\beta$
without loss of generality.
The explicit form of $C(l-\sum_k n_in_{i_k})$ includes multiple products
of the occupation variables, and depends on the parameter $l$ and the lattice
structure.
The free energy density $f$ and average particle density $\rho$ are computed
from the partition function as
\begin{equation}
 f(\mu,L) = -\frac{1}{N}\ln Z(\mu, L),
\end{equation}
and
\begin{equation}
 \rho(\mu,L) = \frac{1}{N}\frac{\partial \ln Z(\mu, L)}{\partial \mu},
  \label{eqn:rho}
\end{equation}
respectively.

\begin{figure}
 \centering
 \includegraphics[width=0.4\linewidth]{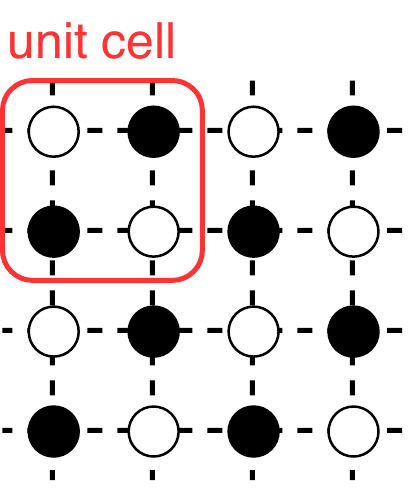}
 \hspace*{0.05\linewidth}
 \includegraphics[width=0.4\linewidth]{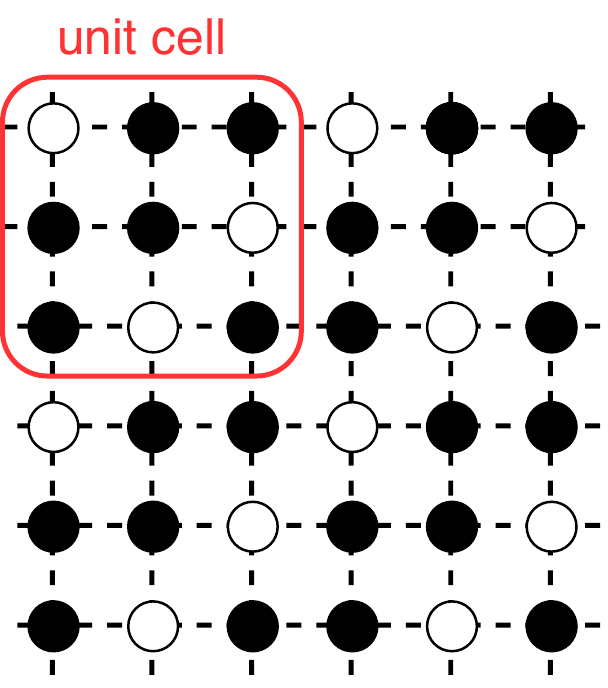}
 \caption{Close-packed structures of the BM model on square lattice with
 $l=0$ (left) and $l=2$ (right). }
 \label{fig:close-pack}
\end{figure}

Thermodynamic functions have a singularity at a transition point.
In the large $\mu$ limit, the particles of the BM model form highly
ordered close-packed structures, as shown in
Fig.~\ref{fig:close-pack} for $l=0$ and $l=2$.
The unit cell of the close-packed structure is $2\times 2$ for $l=0$ and
$3\times 3$ for $l=2$. Correspondingly, the value of the volume fraction
of the close packing is $1/2$ for $l=0$ and $2/3$ for $l=2$.
For $l=0$, the model has a particle-hole symmetry and two-fold degeneracy
in the close-packed structure.
This implies that the transition, if any, is expected to be of second-order and belongs to the 2D Ising universality class.
The BM model with $l=0$ is known as the hard-square lattice gas
model, and its transition point and critical exponent are estimated by the calculation of a corner transfer matrix method.\cite{Baxter1980}
For $l=2$, as shown in Fig.~\ref{fig:close-pack}, the close-packed
structure has a six-fold degeneracy concerning rotation and
translation. A naive argument from the analogy of the 2D Potts
model suggests that the transition is first order.
However, to the best of our knowledge, no previous study has examined
the model with $l=2$.

\section{Method}
\label{sec:method}
Many classical lattice models with local interactions in statistical
physics are expressed by a TN.  The statistical weight for a configuration is
given by a TN, and the partition function is then obtained by
taking the trace of the TN. The HOTRG method\cite{Xie2014}
provides an efficient way to calculate this trace. In this section, we first
discuss how to represent the BM model, including many-body interactions by using the TN, and then explain how to calculate the TN by using the HOTRG.

\subsection{Tensor network representation of the BM model}
\begin{figure}
 \centering
 \includegraphics[width=\linewidth]{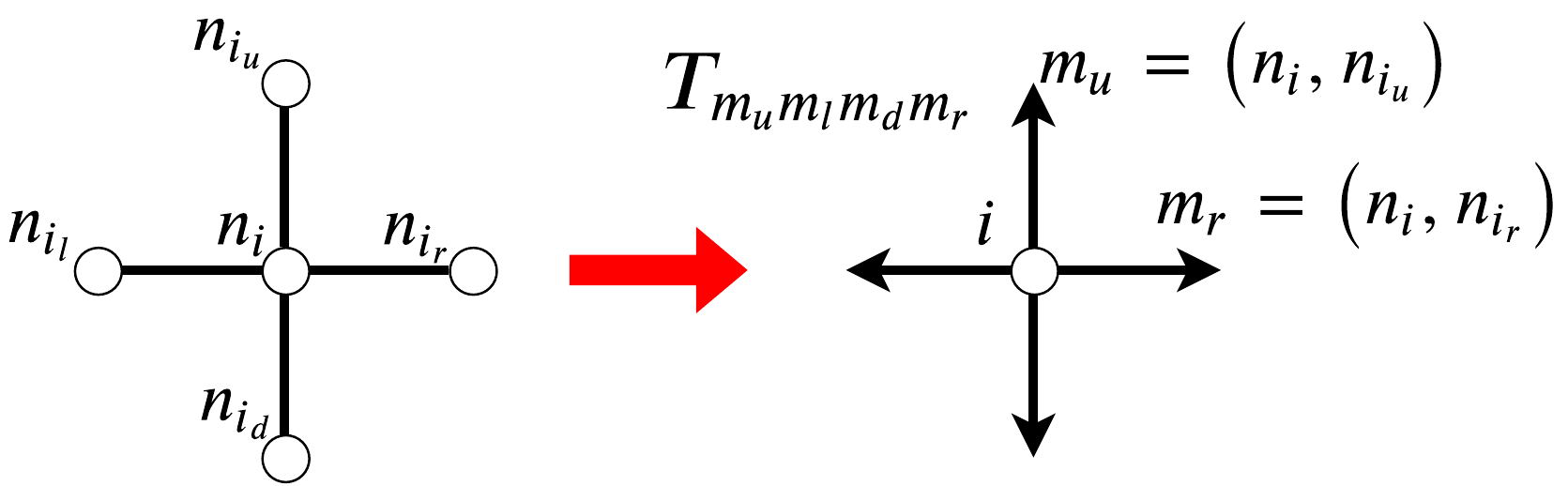}
 \caption{Pair variables representing an occupation configuration of site $i$ and its nearest neighboring sites. The directions of the arrows show the order of the indices. }
\label{fig:tenT}
\end{figure}

We define a tensor $T^{(0)}$ representing a local Gibbs factor determined on a
site with the constraint of the BM model. The superscript
$0$ represents an initial tensor before renormalization.
The initial tensor
$T^{(0)}_{m_um_lm_dm_r}$ at site $i$ has four indices, each of which
is given by $m_k=(n_i,n_{i_k})$, representing pairs of occupation
variables, where the subscript $k\in (u,l,d,r)$, as shown in
Fig.~\ref{fig:tenT}.
Note that the indices of tensor $T^{(0)}$ share the common center site $i$.
It is convenient to define two functions $I_1(m)$ and $I_2(m)$ that
return the first and second index of $m$, respectively.
For example, $I_1(m_u) = n_i$ and $I_2(m_u) = n_{i_u}$.
Then, a consistent configuration of the pair variables $(m_u, m_l, m_d, m_r)$ satisfies the condition:
\begin{equation}
 I_1(m_u)=I_1(m_l)=I_1(m_d)=I_1(m_r).
  \label{eqn:Consistency}
\end{equation}
The hard constraint of Eq.~(\ref{eqn:BMconstraint}) is also expressed using these functions as
\begin{equation}
  l \geq \sum_k I_1(m_k) I_2(m_k) ~\left(=\sum_k n_i n_{i_k}\right)
  \label{eqn:BMconstraint2}
\end{equation}

\begin{figure}
 \centering
 \includegraphics[width=\linewidth]{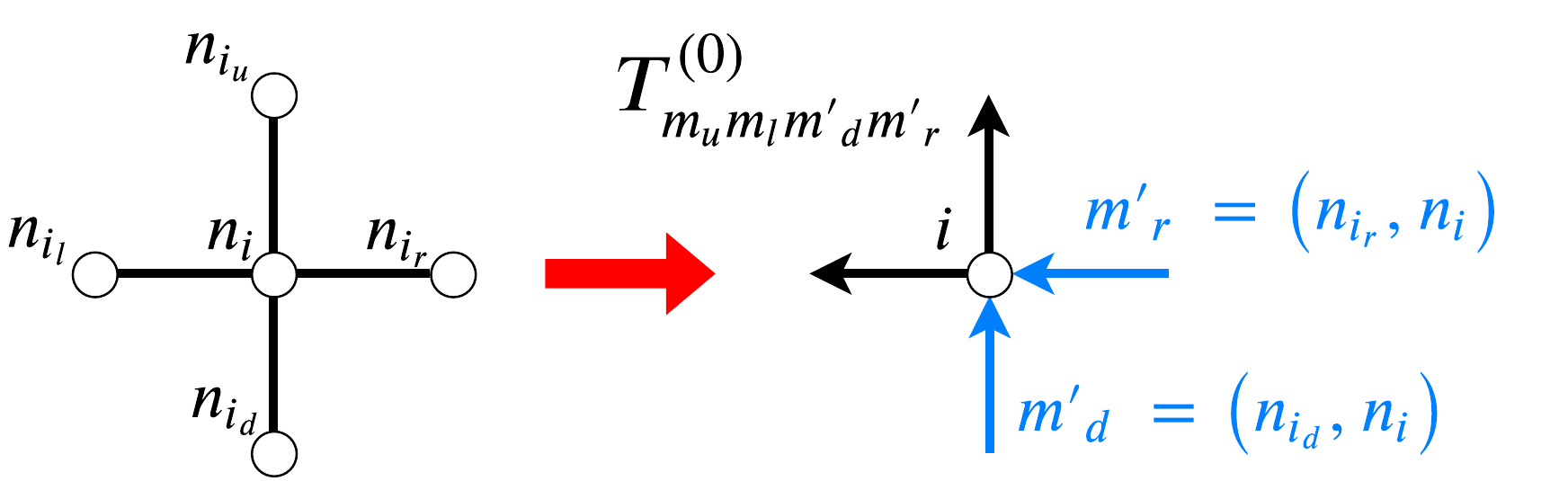}
 \caption{Graphical representation of redefined tensor $T^{(0)}$. The outward
 arrows represent the normal order of pair variables and the inward arrows
 represent the reverse order.}
 \label{fig:tenTd}
\end{figure}

\begin{figure}
 \centering
 \includegraphics[width=\linewidth]{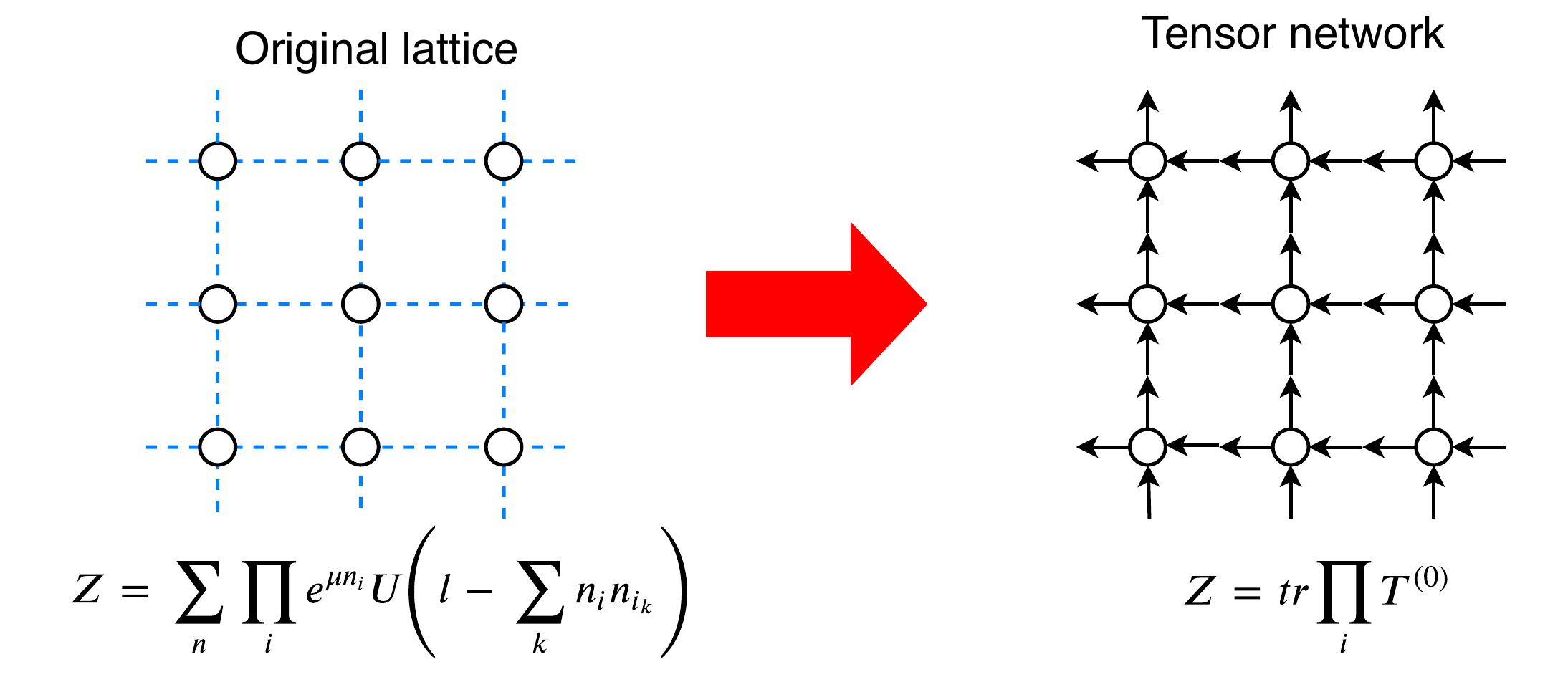}
 \caption{A tensor lattice formed by arranging tensor $T^{(0)}$ on a square lattice. The alignment of the arrows indicates that the problem of mismatch has been solved.}
 \label{fig:Tdlattice}
\end{figure}

A naive definition of the tensor $T$ is given by
\begin{equation}
T_{m_um_lm_dm_r}^{(0)} =
\begin{cases}
 e^{ \mu I_1(m_u)} (= e^{ \mu n_i}), & \mbox{if Eqs. (\ref{eqn:Consistency})
 and (\ref{eqn:BMconstraint2}) are satisfied,} \\
 0&\mbox{otherwise.}
 \label{eqn:formerT}
\end{cases}
\end{equation}
However, we must fix a subtle problem in order to correctly obtain the partition function from the trace of the TN of $T^{(0)}$.
The problem is that the  shared indices of two successive tensors make mismatches in the above definition of the tensor.
Therefore, the lower and right indices, $m_l$ and $m_r$,  of tensor $T^{(0)}$
are redefined to alternative indices, $m'_l$ and $m'_r$, which represent pair variables
$(n_{i_k},n_i)$ with $k \in (d,r)$ in the reverse order from $m_k$.
This definition of indices is shown in Fig.~\ref{fig:tenTd}.
When we array $T^{(0)}$ defined in such a way, the normal-order and reverse-order indices are alternately arranged, and the problem of mismatch described above is solved (Fig.~\ref{fig:Tdlattice}).

Corresponding to the condition of Eq.~(\ref{eqn:Consistency}), the pair variables of redefined $T^{(0)}$ must fulfill this condition:
\begin{equation}
 I_1(m_u) = I_1(m_l) =I_2(m'_d) = I_2(m'_r),
  \label{eqn:Consistency2}
\end{equation}
together with the hard-constraint condition
(\ref{eqn:BMconstraint2}).
The tensor elements of $T^{(0)}$ are given by
\begin{equation}
T^{(0)}_{m_um_lm'_dm'_r} =
 \begin{cases}
 e^{ \mu I_1(m_u)} (=e^{ \mu n_i})& \mbox{if Eqs. (\ref{eqn:Consistency2}) and
 (\ref{eqn:BMconstraint2}) are satisfied, }\\
 0&\mbox{otherwise}.
 \end{cases}
 \label{eqn:properT}
 \end{equation}
Consequently, the partition function of the whole system is expressed by the trace of this TN as
\begin{equation}
  Z = \mathrm{tr} \displaystyle\prod_i^N T^{(0)},
  \label{eqn:BMtensornet}
\end{equation}
where the hard constraints of the BM model in Eq.~(\ref{eqn:BMmodel}) are
fully contained in the expression of the tensor products.

\subsection{Higher order tensor renormalization group}
\begin{figure}
 \centering
 \includegraphics[width=\linewidth]{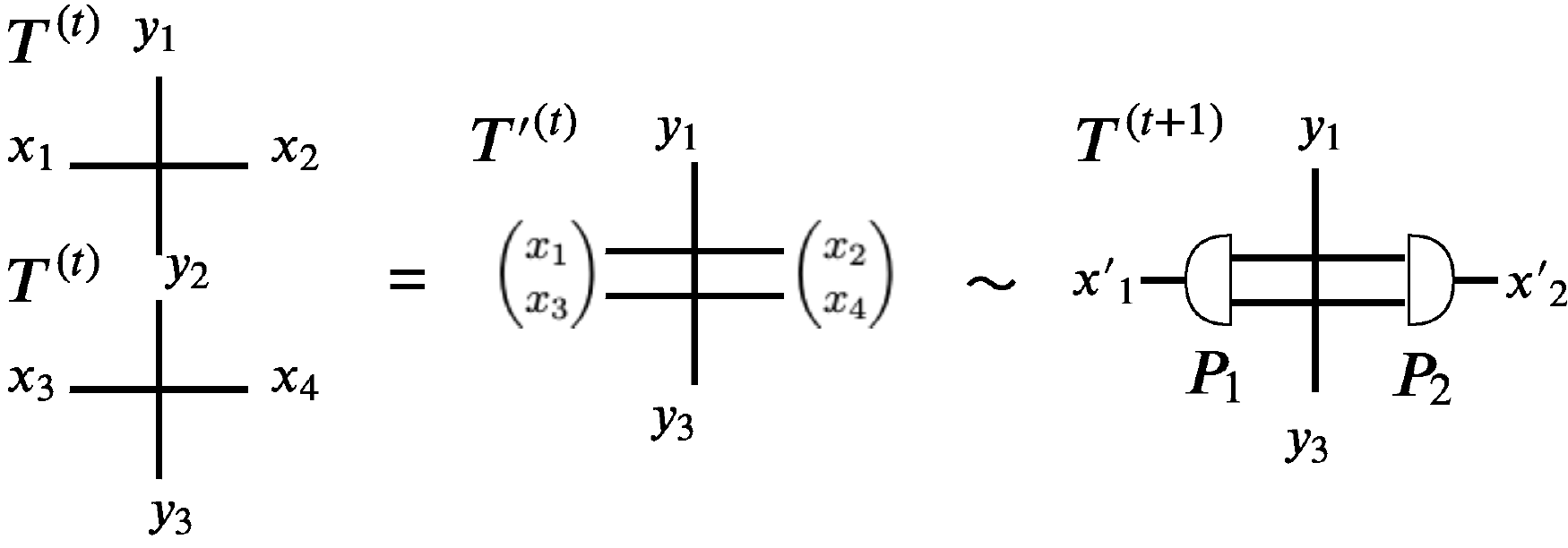}
 \caption{Graphical representation of one renormalization step along the y-axis in the HOTRG method.}
 \label{fig:HOTRG}
\end{figure}

Once representing the partition function as a trace of the TN, we calculate it using the HOTRG method\cite{Xie2012}, which is a variant of TRG methods.
In this method, the TN is contracted sequentially along the $x$- and $y$- axes alternately.
The tensor after $t$ steps of renormalization is denoted by $T^{(t)}$.
Fig.~\ref{fig:HOTRG} shows the $(t+1)$-th step of renormalization by
contraction along the $y$-direction.
We first consider the contraction of successive tensors as
\begin{equation}
  T'^{(t)}_{y_1,x_1x_3,y_2,x_2x_4}=
   \sum_{y_2}T^{(t)}_{y_1,x_1,y_2,x_2}T^{(t)}_{y_2,x_3,y_3,x_4}.
  \label{eqn:contraction}
\end{equation}
The dimension along the $x$ direction of tensor $T'^{(t)}$ is increased by
repeating the contraction, which renders the computational cost increase exponentially.
The upper limit of this dimension is reduced to a suitable constant
$D_{\text{cut}}$, often called bond dimension, by using projection tensors $P^{(t)}_1$ and $P^{(t)}_2$,
which are chosen to maintain the value of the TN as much as
possible (details are provided in Appendix~\ref{sec:projectors}).
Consequently, the $(t+1)$-th tensor $T^{(t+1)}$ is obtained by
\begin{equation}
  T^{(t+1)}_{y_1,x'_1,y_2,x'_2}= \sum_{x_1\sim
   x_4}T'^{(t)}_{y_1,x_1x_3,y_2,x_2x_4}{P_1^{(t)}}_{x_1x_3,x'_1}{P_2^{(t)}}_{x'_2,x_2x_4}.
  \label{eqn:HOTRG}
\end{equation}
The above two operations of Eqs.~(\ref{eqn:contraction}) and (\ref{eqn:HOTRG}) are simply represented as
\begin{equation}
  T^{(t+1)}\leftarrow T^{(t)}T^{(t)}.
  \label{eqn:renormalize}
\end{equation}
$T^{(t+2)}$ is then calculated by renormalizing $T^{(t+1)}$ along the $x$-axes in the same way.
A large square lattice is computed by alternately renormalizing along the $y$- and $x$-axes.
The trace of the renormalized tensor $T^{(t)}$
gives an approximation of the partition function under the periodic
boundary conditions of the system of size $N=2^t$, described by
\begin{equation}
  Z \simeq \mathrm{tr} T^{(t)}\equiv \sum_{a,b}T^{(t)}_{abab}.
\end{equation}

\subsection{Impurity tensor}
The partition function can be calculated by the HOTRG method described
above, and the average particle density $\rho$ of the system can be
calculated by numerically differentiating its logarithm as in Eq.~(\ref{eqn:rho}) in principle.
However, in general, numerical differentiation involves large numerical errors.
The impurity tensor\cite{Gu2008,Morita2019} provides us a useful method to calculate physical quantities using a TN without numerical differentiation.
This method is based on the fact that physical quantities can be expressed by
a ratio of the values of the TNs with and without impurity tensors.

First, we consider the average particle density, which is the ensemble average of the occupation number defined on each site.
A tensor $S_1^{(0)}$, called an
impurity tensor, is defined at site $i$ as a product of a local physical quantity $n_i$ and the original tensor $T^{(0)}(i)$,
\begin{equation}
  S_1^{(0)}(i) = n_iT^{(0)}(i).
  \label{eqn:impurity}
\end{equation}
where the argument $i$ indicates where the tensor is located.

Here, the partition function of Eq.~(\ref{eqn:BMtensornet}) is formally rewritten as
\begin{equation}
  Z = \text{tr}\left[T^{(0)}(0)T^{(0)}(1)\ldots T^{(0)}(N)\right].
\end{equation}
We also consider another TN where only the tensor at site $i$ is replaced by the impurity tensor, defined as
\begin{equation}
  Z_i = \mathrm{tr}\left[T^{(0)}(0)T^{(t)}(1)\ldots T^{(0)}(i-1)~S_1^{(0)}(i)~T^{(0)}(i+1)\ldots T^{(0)}(N)\right].
\end{equation}
The ratio of the trace of these TNs gives the ensemble average $\langle
n_i\rangle$ of the local occupation number $n_i$ at site $i$ as \begin{equation}
  \left<n_i\right> = \frac{Z_i}{Z}
  \label{eqn:local}
\end{equation}
The average particle density $\rho$ is thus calculated by
\begin{equation}
  \rho =\left<n \right> \equiv \left<\frac{1}{N}\sum_i n_i \right> = \frac{1}{N}\sum_i\left<
	  n_i \right>.
  \label{eqn:impurity net2}
\end{equation}
Corresponding to steps of the renormalization of Eqs.~(\ref{eqn:contraction}) and (\ref{eqn:HOTRG}) for the tensor $T^{(t)}$,
the impurity tensor is renormalized by the following two steps:
\begin{eqnarray}
  {{S'_1}^{(t)}}_{y_1,x_1x_3,y_2,x_2x_4} & = &
   \frac{1}{2}\sum_{y_2}\left({S_1^{(t)}}_{y_1,x_1,y_2,x_2}
			 T^{(t)}_{y_2,x_3,y_3,x_4} \right. \ \ \ \ \nonumber  \\
&&		\left.	 +T^{(t)}_{y_1,x_1,y_2,x_2}{S_1^{(t)}}_{y_2,x_3,y_3,x_4}\right),
  \label{eqn:contraction1}
\end{eqnarray}
and
\begin{equation}
  {S_1^{(t+1)}}_{y_1,x'_1,y_2,x'_2}= \sum_{x_1\sim
   x_4}{{S'_1}^{(t)}}_{y_1,x_1x_3,y_2,x_2x_4}{P_1^{(t)}}_{x_1x_3,x'_1}{P_2^{(t)}}_{x'_2,x_2x_4},
  \label{eqn:HOTRG1}
\end{equation}
where in the former the impurity is locally averaged and in the latter
the projection tensors $P_1^{(t)}$ and $P_2^{(t)}$ are the ones used in Eq.~(\ref{eqn:HOTRG}). This renormalization step can be expressed as follows, similar to Eq.~(\ref{eqn:renormalize}):
\begin{equation}
  S_1^{(t+1)}\leftarrow \frac{1}{2}\left(S_1^{(t)}T^{(t)} +T^{(t)}S_1^{(t)} \right)
  \label{eqn:renormalize1}
\end{equation}
Then, the average particle density $\rho$ for the system with size $N=2^t$ is
calculated by
\begin{equation}
  \rho \simeq \frac{1}{Z}\text{tr} S_1^{(t)}.
  \label{eqn:rho_impurity}
\end{equation}

Similarly, to calculate the $k$-th moment of the mean of the
occupation number $\left<n^k\right>$ by the impurity tensor method,
we define the $k$-th impurity tensor $S_k^{(0)}$ with the $k$-th power of the occupation number in a tensor as
\begin{equation}
  S_k^{(0)}(i) = n_i^kT^{(0)}.
  \label{eqn:impurity_k}
\end{equation}
Considering how the $k$ impurities are contained in the TN, a renormalization step of the $k$-th impurity tensor $S_k$ is expressed as
\begin{equation}
  \displaystyle S_k^{(t+1)}\leftarrow \frac{1}{2^k}\left(S_k^{(t)}T^{(t)} +\sum_{i=1}^{k-1} \binom{k}{i}~S_{k-i}^{(t)}S_{i}^{(t)} +T^{(t)}S_k^{(t)} \right).
  \label{eqn:renormalize_k}
\end{equation}
From these impurity tensors, the $k$-th moment $\left<n^k\right>$ of
the system of size $N=2^t$, as in Eq.~(\ref{eqn:rho_impurity}), is
calculated by
\begin{equation}
  \left<n^k\right> = \frac{1}{Z}\text{tr} S_k^{(t)}.
\end{equation}

For the case of the BM model with $l=0$, the close-packed structure is the
configuration in which either sublattice is completely occupied.
The order parameter $m$ is described by

\begin{equation}
  m \equiv \frac{1}{N}\left(\sum_{i\in\Lambda_a} n_i -
		       \sum_{i\in\Lambda_b} n_i\right),
  \label{eqn:order parameter}
\end{equation}
where $\Lambda_a$ and $\Lambda_b$ denote two sublattices.
The $k$-th moment $\left<m^k\right>$ can be calculated by changing the
definition of the impurity tensor of Eq.~(\ref{eqn:impurity_k}) to
\begin{equation}
  S_k^{(0)}(i) =
  \begin{cases}
    +n^k_iT^{(0)},   ~~~~~~~~(\text{if} ~~i \in \Lambda_a ) \\
    -n^k_iT^{(0)}.  ~~~~~~~~(\text{if} ~~i \in \Lambda_b )
  \end{cases}
\end{equation}
Using these moments, we also define the Binder parameter of the order
parameter $m$ as
\begin{equation}
  U =
   \frac{1}{2}\left(3-\frac{\left<m^4\right>}{\left<m^2\right>^2}\right),
  \label{eqn:Binder}
\end{equation}
which is useful for determining the transition point for the
second-order transition.

\section{Results}
\label{sec:results}
In this section, we present results obtained by using the HOTRG for the
BM model with $l=0$ and $l=2$, and discuss the numerical accuracy of the HOTRG in comparison with the MCMC method.

\subsection{BM model with $l=0$}
\label{sec:mu_c l=0}
\begin{figure}
 \centering
 \includegraphics[width=\linewidth]{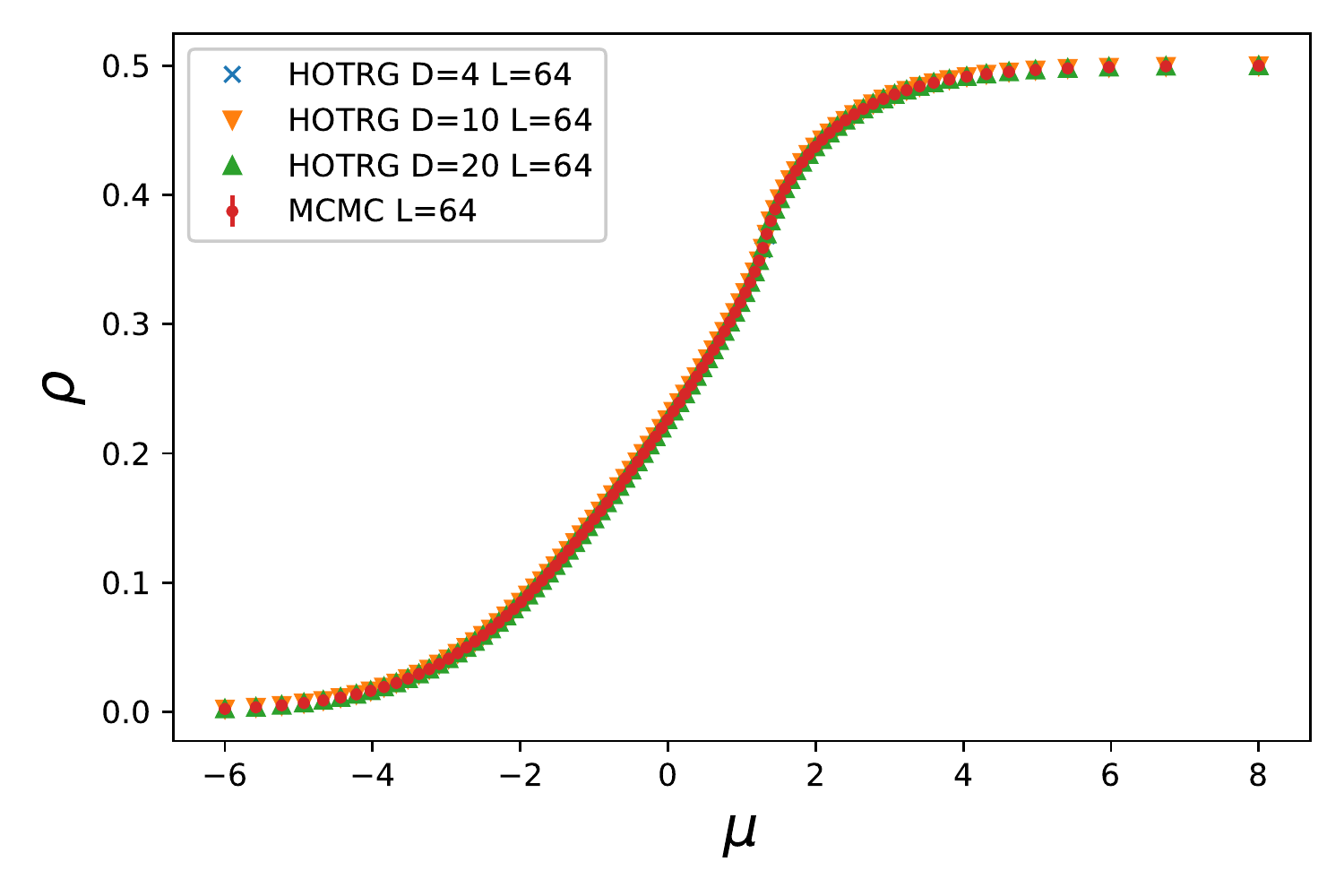}
 \caption{Dependence of $\rho$ on chemical potential for the BM model with
 $l=0$. The system size is $L=64$. Each data is calculated by HOTRG
 with $D_{cut}=4$, $10$,  $20$
 and the MCMC method. }
 \label{fig:l=0}
\end{figure}

Fig.~\ref{fig:l=0} shows average density $\rho$ as a
function of the chemical potential $\mu$ for the BM model with $l=0$.
We compare the HOTRG calculation with the MCMC calculation
obtained by using the exchange MC method.\cite{Hukushima1996}
The HOTRG calculations are performed with different values of the bond
dimension $D_{\rm cut}$.
They agree with each other, even when $D_{\rm cut}=4$.
In the large $\mu$ limit, the density approaches the expected value of
the close packing.
This suggests that small values of $D_{\rm cut}$ are sufficient for the HOTRG calculation of this model.

\begin{figure}
 \centering
 \includegraphics[width=\linewidth]{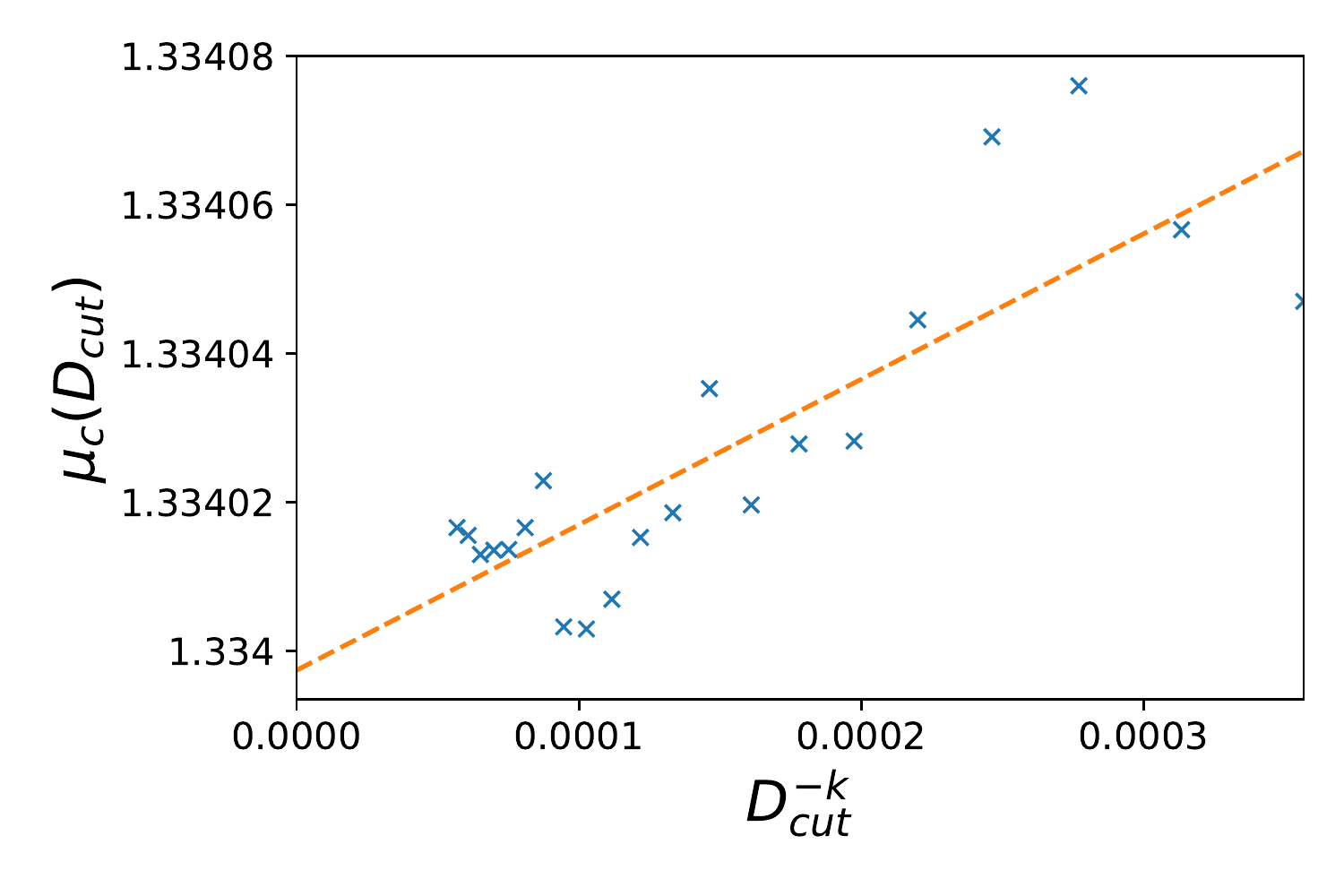}
 \caption{
 An effective transition point $\mu_c(D_\mathrm{cut})$ as a function of
 $D_\mathrm{cut}^{-k}$ with $k=2.633(6)$ for the BM model with $l=0$. Effective transition points are determined by the jump of the Binder parameter calculated by the HOTRG with a finite $D_\mathrm{cut}$.
 The dotted line represents the result obtained by the least-squares
 method. }
 \label{fig:scaling_l=0}
\end{figure}

We also calculate the Binder parameter of Eq.~(\ref{eqn:Binder}) by the HOTRG for large sizes up to $L=2^{25}$, which are challenging by the MCMC
method in equilibrium.
In such a large system, the Binder parameter jumps approximately at the
transition point $\mu_c$, which weakly depends on the bond dimension
$D_{\rm cut}$.
The transition point is estimated by extrapolating the values of
an effective transition point $\mu$ with a finite $D_{\rm cut}$.
Previous studies\cite{Morita2019,Ueda2014} have suggested that the displacement
of the transition point $\Delta\mu_c=|\mu_c(D_{\rm cut})-\mu_c(\infty)|$ follows a power law
\begin{equation}
 \Delta\mu_c\propto D_{\rm cut}^{-k}
  \label{eqn:extrapolation}
\end{equation}
with an exponent $k$.
We estimate effective transition points $\mu_c(D_\mathrm{cut})$ for
each $D_\mathrm{cut}$. As shown in Fig.~\ref{fig:scaling_l=0}, the
least-squares method yields
\begin{equation}
  \mu_c=1.33400(1),
  \label{eqn:mu_c}
\end{equation}
which coincides with the result $\mu_c=1.33400(3)$
by the corner transfer matrix approximation method\cite{Baxter1980}.

\begin{figure}
 \centering
 \includegraphics[width=\linewidth]{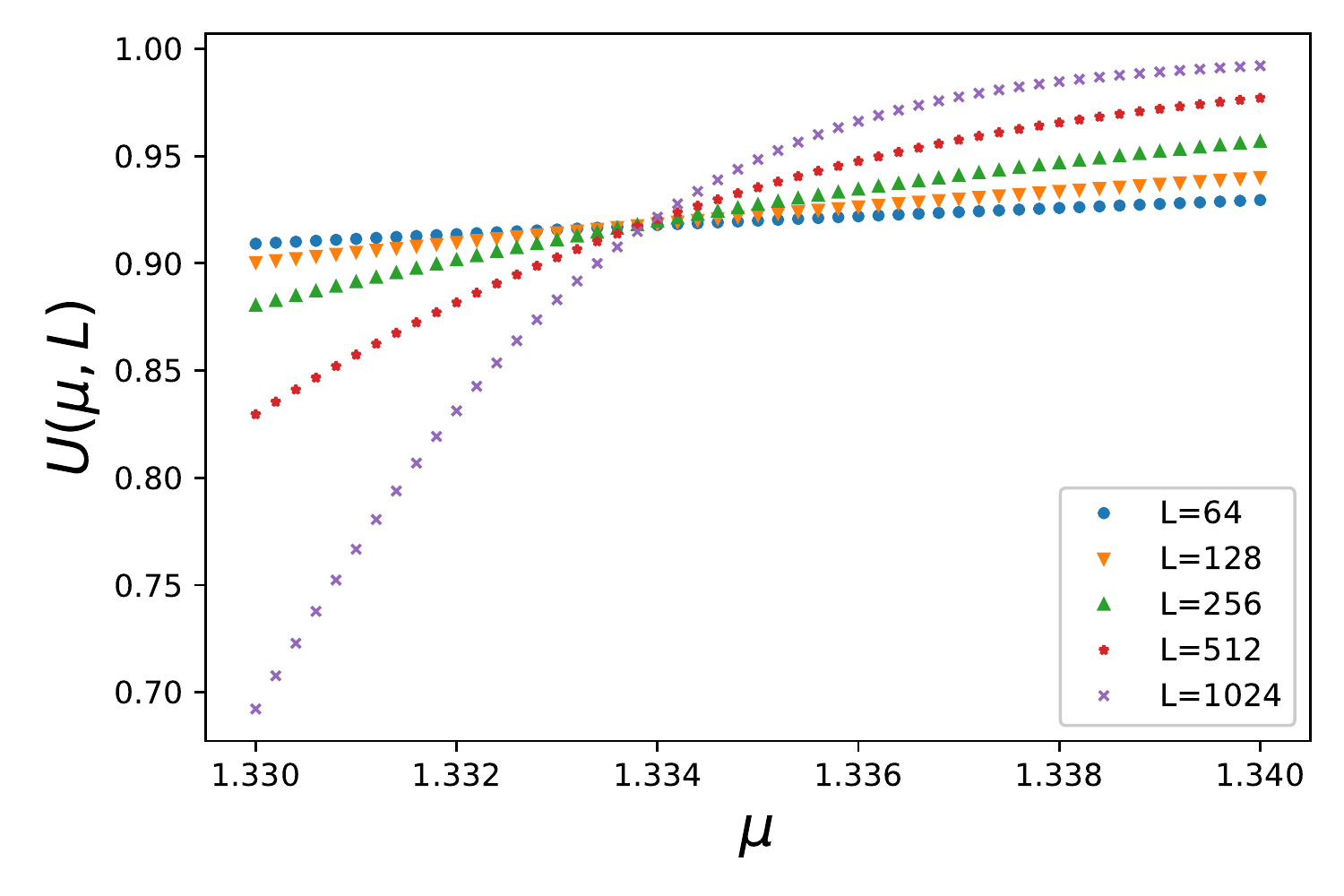}
 \caption{Dependence of the Binder
 parameter $U(\mu,L)$ on chemical potential for different sizes obtained by the HOTRG with $D_{\text{cut}}=40$.}
 \label{fig:Binder}
\end{figure}

\begin{figure}
 \centering
 \includegraphics[width=\linewidth]{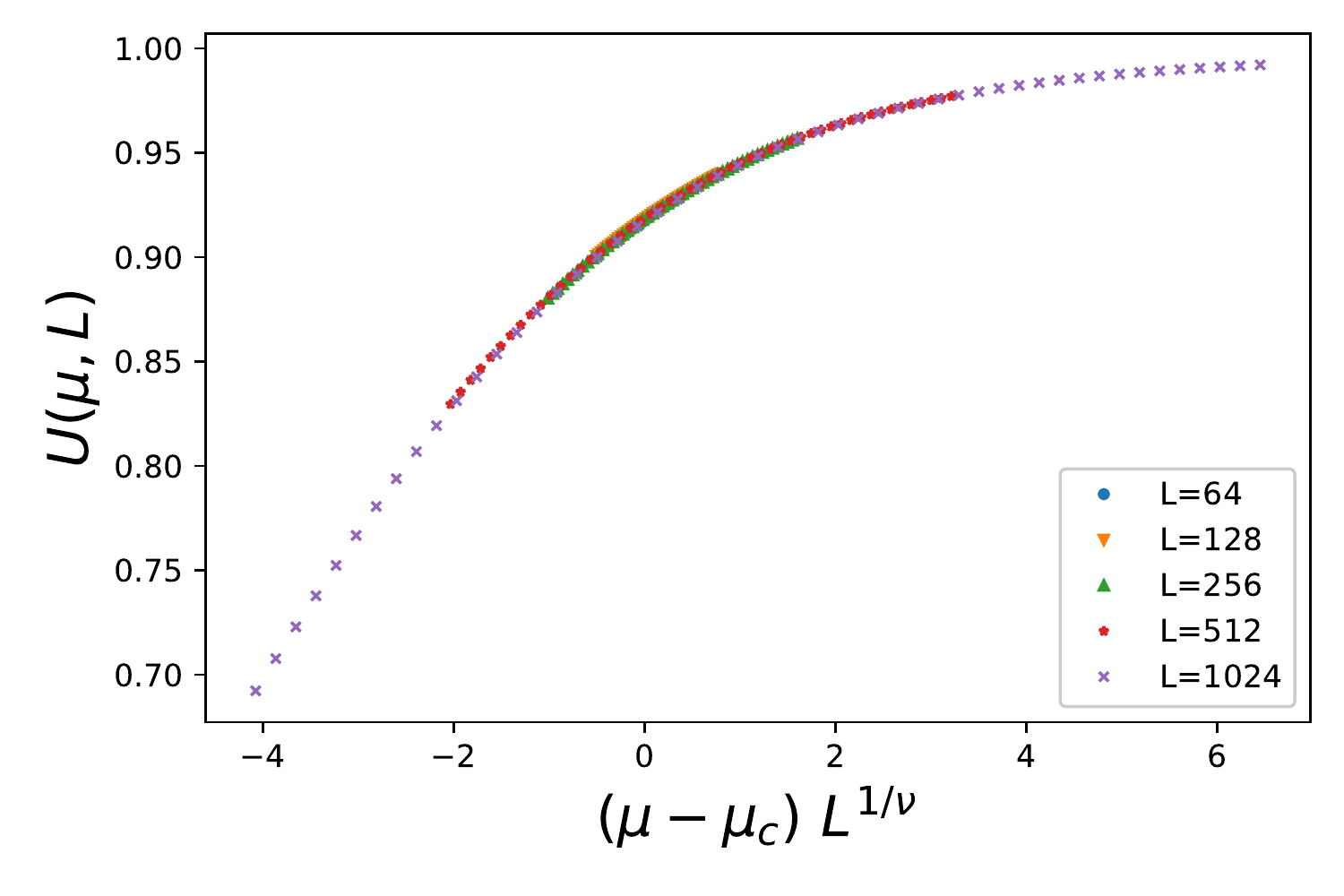}
 \caption{Finite-size scaling plot of the Binder parameter with $\mu_c = 1.33389(6)$ and $\nu = 1.03(3)$. The data used are the same as in Fig.~\ref{fig:Binder}.}
 \label{fig:Binder_FSS}
\end{figure}

In addition to the above analysis, we subsequently investigate the finite-size effect of the Binder parameter $U(\mu, L)$ near the transition point for a sufficiently large $D_{\rm cut}$.
As shown in Fig.~\ref{fig:Binder}, the Binder parameters of different sizes clearly intersect, indicating that the transition is second-order.
To extract the transition point and the correlation-length exponent from these
data, we perform the finite-size scaling analysis using the Bayesian scaling analysis\cite{Harada2011}, which works well as shown in Fig.~\ref{fig:Binder_FSS}. The analysis yields
\begin{equation}
  \mu_c = 1.33389(6)\ \   \mbox{and}~~\nu = 1.03(3),
\end{equation}
which are consistent with the previous estimate of Eq.~(\ref{eqn:mu_c})
and the 2D Ising universality class, respectively.

\subsection{BM model with $l=2$}
 \label{sec:l=2 result}
\begin{figure}
 \centering
 \includegraphics[width=\linewidth]{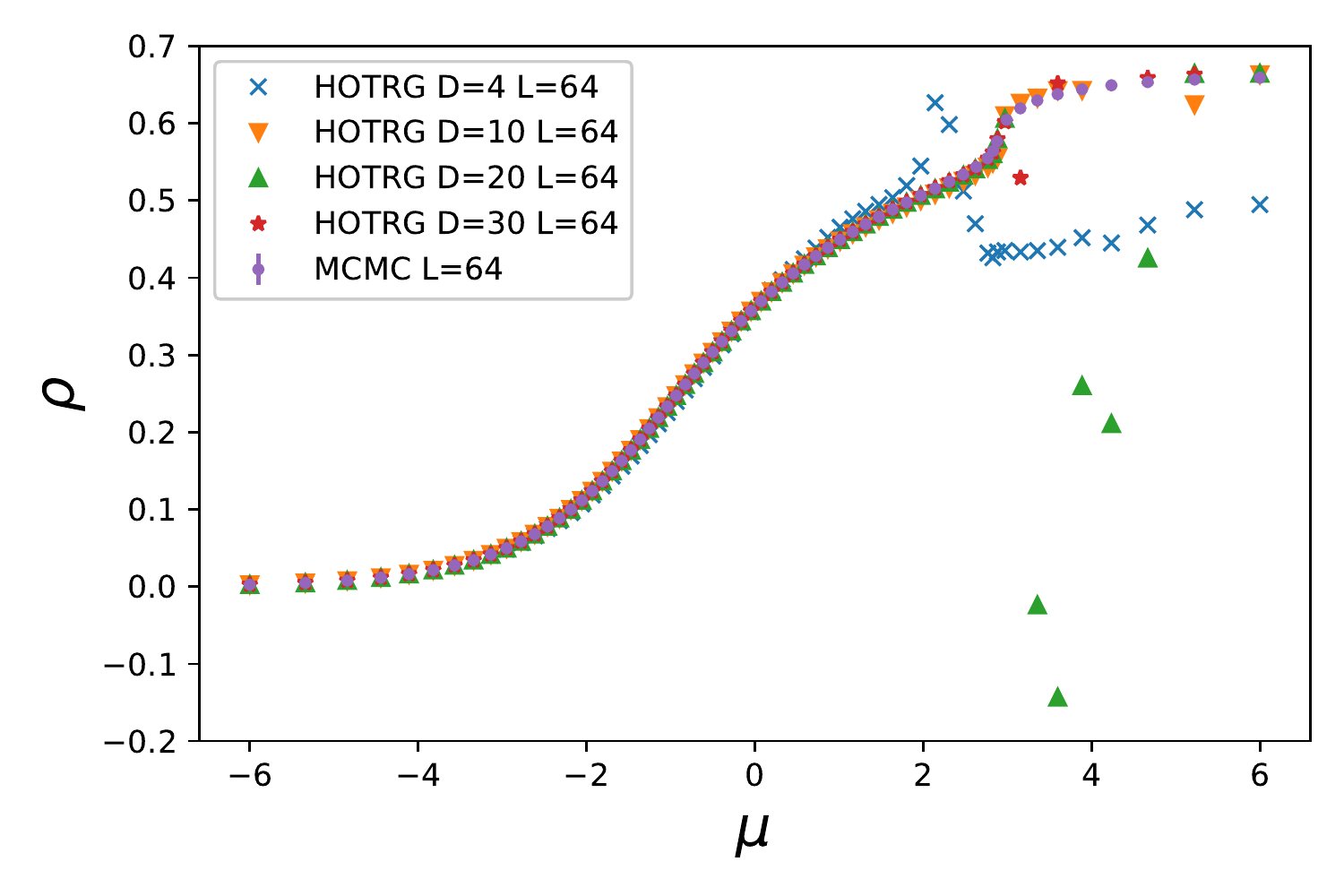}
 \caption{Dependence of $\rho$ on chemical potential for the BM model with
 $l=2$ and $L=64$ calculated by the HOTRG and MCMC methods.}
 \label{fig:l=2_L=64}
\end{figure}

Now, we move on to the application of the HOTRG method performed to the BM
model with $l=2$.
Fig.~\ref{fig:l=2_L=64} shows the dependence of $\rho$ on chemical potential for $L=64$.
The result obtained by the HOTRG does not coincide with that yielded by MCMC, particularly in the high-density region, whereas the results of the two methods are consistent with each other in the low-density region.
The average density depends on $D_\mathrm{cut}$ in a non-systematic way
and it can even take on negative values that are
unphysical.
In contrast to the case with $l=0$, the numerical accuracy of the HOTRG is not
satisfactory even for a large value of $D_\mathrm{cut}$, such as $D_\mathrm{cut}=30$.

This is likely due to a mismatch between the lattice calculated by the HOTRG and the ordered, close-packed structure of the model.
While the unit cell of the close-packed structure for $l=0$ is commensurate with the $L=2^n$ lattice calculated by the HOTRG,
this is not the case for $l=2$.
As a result, the densely packed states for large values of $\mu$ in the $L=2^n$
system include states that are not close-packed states, represented by small
singular values in the HOTRG.
The contributions of such states with small singular values are truncated
in the renormalization procedure of the HOTRG, leading the poor numerical
accuracy in systems with incommensurate states, as shown in Fig.~\ref{fig:l=2_L=64}

\begin{figure}
 \centering
 \includegraphics[width=\linewidth]{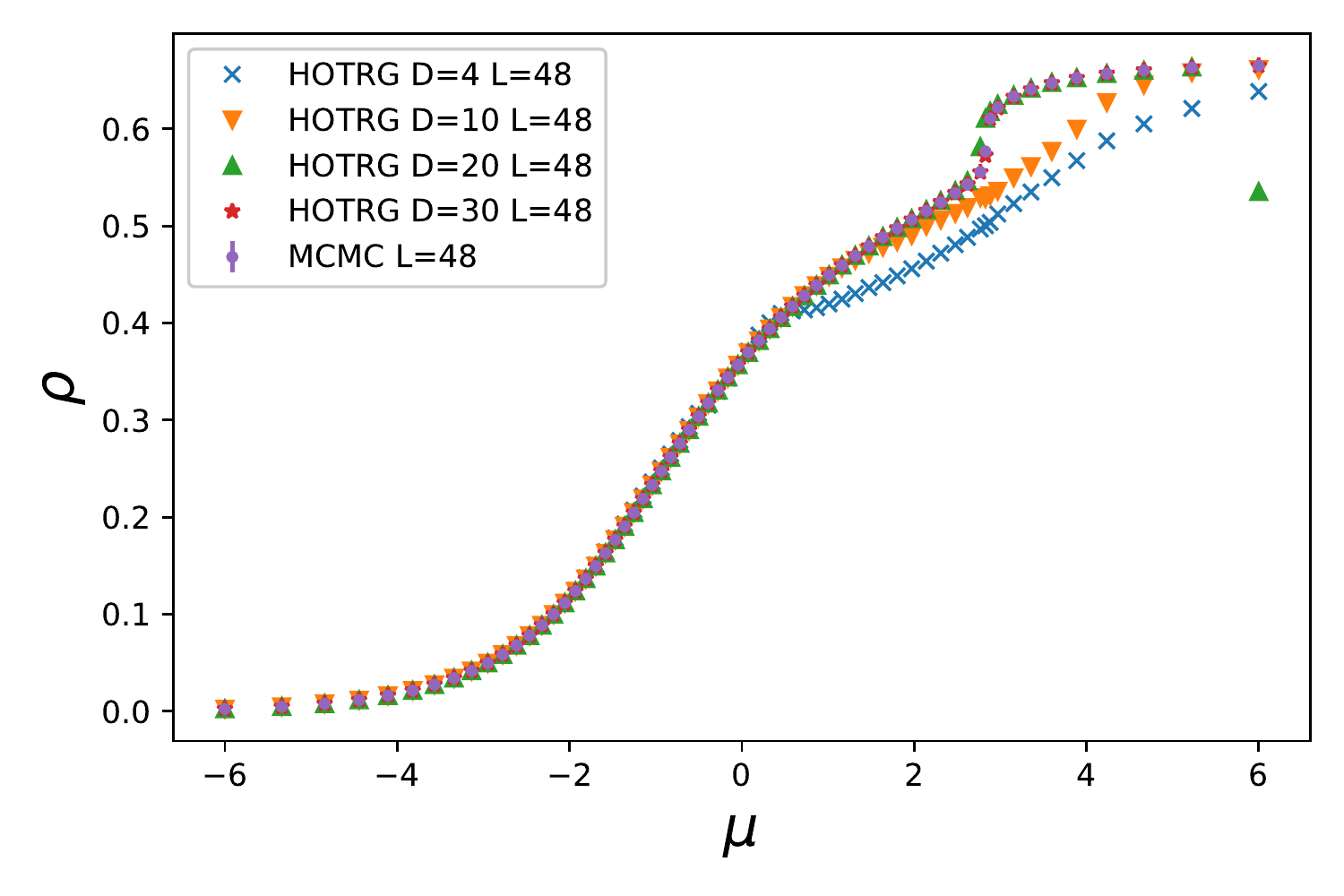}
 \caption{Dependence of $\rho$ on chemical potential for the BM model with
 $l=2$. Each data is calculated by the HOTRG with $D_\text{cut}=4$, $10,20$ and $30$ for the system size $L=48$, and by using the MCMC method.}
 \label{fig:rho_l=2}
\end{figure}

To avoid this mismatch, we examine another lattice unit. As mentioned
in Sec.~\ref{sec:model}, the size of the unit cell of close-packed structure is $3\times3$ for the case $l=2$.
Therefore, starting from a bundle of $3\times3$ original initial tensors, HOTRG calculations are performed for systems of size $L=3\times2^n$, which are commensurate with the close-packed structure.
Fig.~\ref{fig:rho_l=2} shows that the calculation using the HOTRG for size $L=3\times 2^4$ and a sufficiently large $D_\mathrm{cut}$ is consistent with the results of the MCMC.

\begin{figure}
 \centering
 \includegraphics[width=\linewidth]{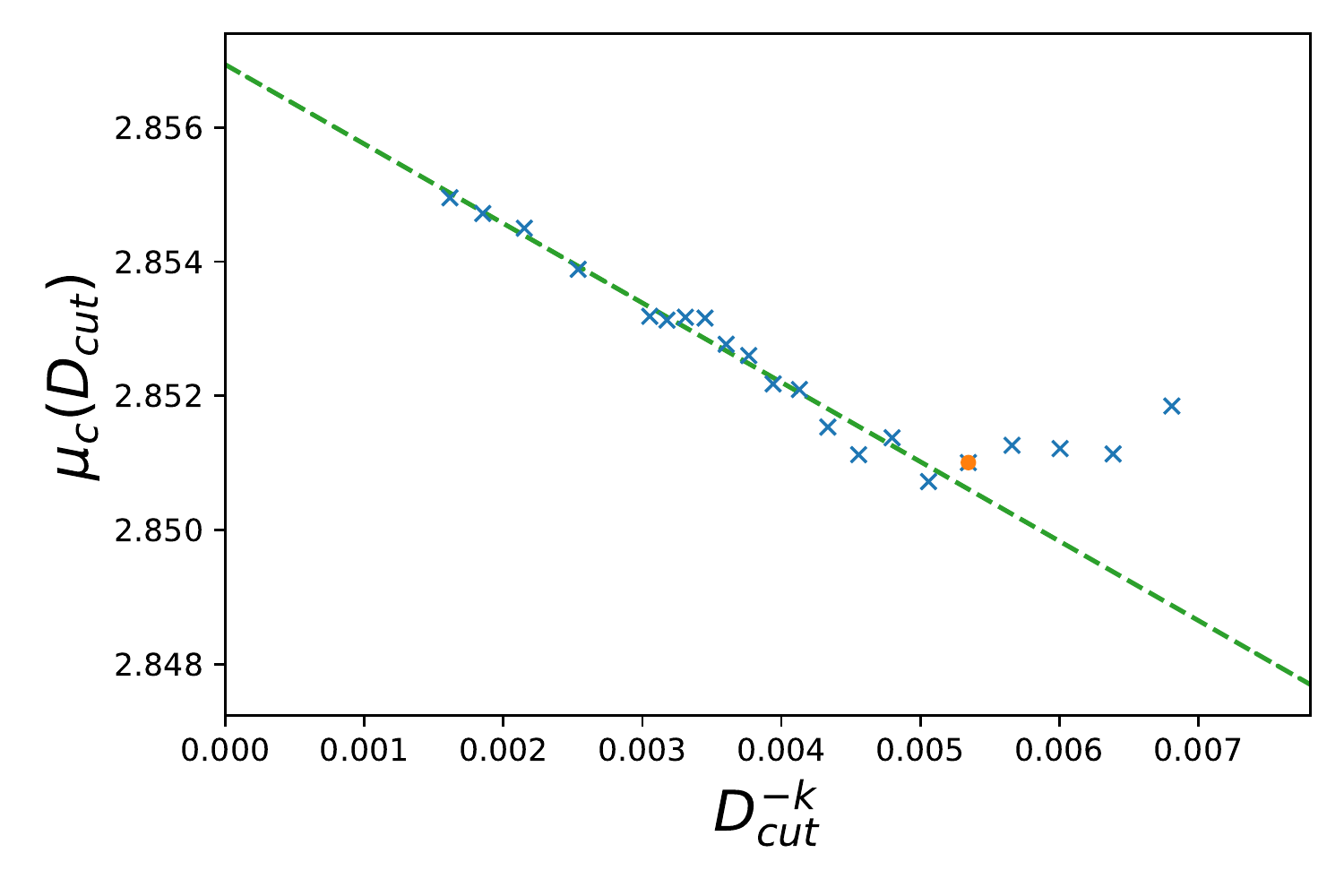}
 \caption{The effective transition point $\mu_c$ as a function of
 $D_\mathrm{cut}^{-k}$ with $k = 1.57(1)$ for the BM model with $l=2$. The
 values of $\mu_c(D_\mathrm{cut})$ are determined by the jump in the average
 density calculated by the HOTRG with a finite $D_\mathrm{cut}$. The dotted
 line represents the result obtained by the least-squares
 method for the date of $D_\mathrm{cut}>28$.}
 \label{fig:scaling_l=2}
\end{figure}

As discussed in Sec.~\ref{sec:model}, the BM model with $l=2$ is
expected to have a first-order transition. Then, the average density
would jump at the transition point in the thermodynamic limit, as with the
Binder parameter for the second-order transition discussed in the
previous subsection.
To determine the transition point, we calculate the average density $\rho$ of the very large size, $L=3\times 2^{20}$.
We estimate an effective transition point $\mu_c(D_\mathrm{cut})$ at which $\rho$ jumps depending on $D_\mathrm{cut}$.
The transition point is estimated as $\mu_c=2.8569(4)$ by the least-squares method to Eq.~(\ref{eqn:extrapolation}) using the date of $D_\mathrm{cut}>28$.
As shown in Fig.~\ref{fig:scaling_l=2},
the data with $D_\mathrm{cut}\leq 28$ are clearly deviated from the
regression line.
This suggests that extrapolation of $D_\mathrm{cut}$ to infinity requires
caution in assuming the simple formula of Eq.~(\ref{eqn:extrapolation})
as an asymptotic form.

\section{Summary and Discussion}
\label{sec:discussions}
This study developed a TN for the 2D BM model that has constraints expressed by local many-body interactions.
Using the TN, we applied the HOTRG method to the BM model that exhibits first- and second-order transitions to the dense crystal phase depending on the model parameter.
The HOTRG method enabled us to accurately estimate the transition points
for the first- and second-order transitions by calculating for much
larger sizes that can be achieved using the MCMC method.

We have also found that the numerical accuracy of the calculation of the partition function using the HOTRG method can be extremely poor, and yields negative values.
This is because the densely packed states of the model are not commensurate
with the lattice structure of the TN.
This problem can be partially solved by matching the lattice unit of the
initial tensor to that of the densely packed states, as in the BM model
with $l=2$ discussed in the previous section.

However, this method has a disadvantage because it requires knowing the
close-packed structure a prior, and the size of the unit cell needs to be smaller than that of the initial tensor that can be calculated by HOTRG.
The BM model with $l=1$ on a honeycomb lattice, for which the close-packed structure has not been elucidated and the unit cell of a possible dense state
consists of 13 hexagons, is an example that meets this
difficulty\cite{Hukushima2010}.
As a practical solution, we consider the following decomposition of the
partition function $Z\simeq\text{tr}T=\sum_{a,b}T_{abab}$:
\begin{equation}
 Z \simeq
  Z_{\rm norm} \times Z_{\rm direction}
  \label{eqn:norm_direction},
\end{equation}
where
$Z_{\rm norm} = \sqrt{\sum_{abcd} T^2_{abcd}}$ and $Z_{\rm
direction}=\text{tr}T/Z_{\rm norm}$.\\
Since $Z_\text{norm} = \mathcal{O}(e^N)$ and $Z_\text{direction} = \mathcal{O}(1)$ for sufficiently large $N$,
the contribution of $Z_{\rm direction}$ to the free-energy density is
negligibly small.
The contribution of the lattice mismatch to the free-energy density is
expected to be of the order of the surface term, and the physical quantities can be estimated approximately from the principal term $Z_{\rm norm}$ alone.

\begin{figure}
 \centering
 \includegraphics[width=\linewidth]{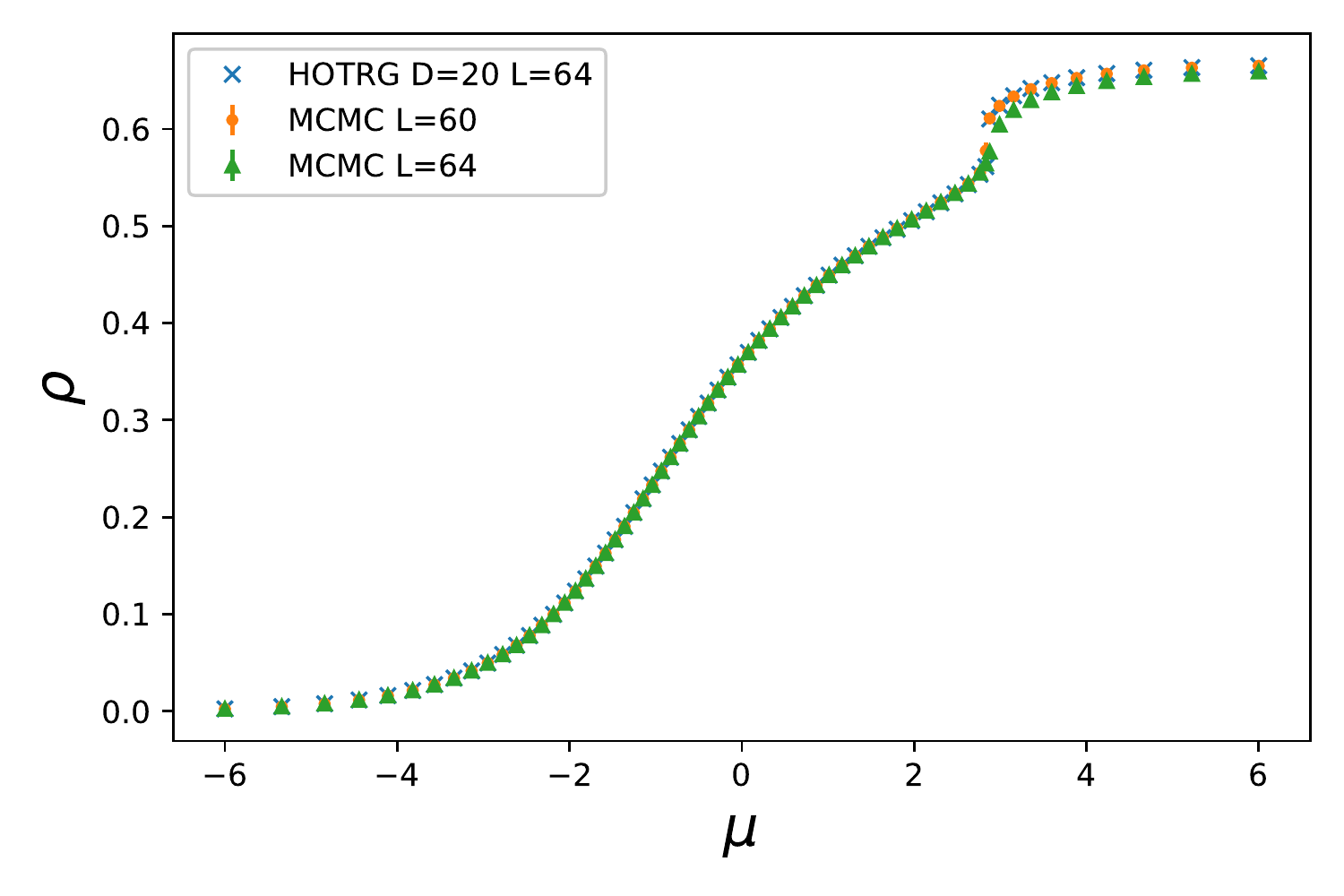}
 \caption{Dependence of the average density $\rho$ on chemical potential for the BM model with $l=2$. The results of the HOTRG method are marked by crosses and the results of the MCMC method are by filled circles.}
 \label{fig:no_norm}
\end{figure}

Fig.~\ref{fig:no_norm} shows the density $\rho$ obtained by
numerically differentiating the principal term $Z_{\rm norm}$ through
Eq.~(\ref{eqn:rho}) of the BM model with $l=2$ for $L=64$. The MCMC
results are calculated for the sizes of $L=60$ and $64$, in which the
latter, which is incommensurate with the close-packed structure, has a
slightly smaller value in the high-density region than the former.
The calculation of HOTRG with $L = 64$ is consistent with the MCMC
result of $L = 60$ and deviates from that of $L=64$.
This is because HOTRG using $Z_{\rm norm}$ only is a calculation for the thermodynamic limit and does not take into account surface effects correctly.
Therefore, this method is not suitable when an accuracy of $O(1/N)$ is
needed to measure physical quantities. Moreover, the impurity tensor method cannot
be applied because the calculation of the partition function ratio requires
the accuracy of the partition functions itself.

In summary, this study has demonstrated that TRG methods can be used for lattice glass models with local constraints expressed by many-body interactions, and this is not restricted to ordinary systems that contain only two-body interactions.
Although the present model studied in this paper did not show a glass
transition, we believe that this can provide a new numerical method for studying the glassy systems from the perspective of equilibrium statistical mechanics.

\section*{Acknowledgments}
The authors thank Satoshi Morita and Yoshihiko Nishikawa for many useful
discussions.
They also thank Jun Takahashi for the careful reading of the manuscript.
This work was supported by a Grant-in-Aid from JSPS KAKENHI, Grant Numbers 17H02923, and 19H04125.

\appendix
\section{How to calculate the projectors in HOTRG}
\label{sec:projectors}
In this appendix, we explain how to construct the projection tensors
$P^{(t)}_1$ and $P^{(t)}_2$ used in our work \cite{Iino2019}.
In the calculation of HOTRG, the projection tensors are used for dimensional compression to preserve the value of the TN as much as possible.  They are obtained by minimizing the error function defined as
\begin{equation}
  \epsilon=\left|\left|M_1^{(t)}P^{(t)}_1{P^{(t)}_2}{M_2^{(t)}}^{\dagger}-M_1^{(t)}{M_2^{(t)}}^{\dagger}\right|\right|^2,
  \label{eqn:epsilon}
\end{equation}
where
\begin{equation}
{M_1^{(t)}}_{x_1x_3y_1y_2,x_2x_4}={M_2^{(t)}}^{\dagger}_{x_1x_3,y_1y_2x_2x_4}=T'^{(t)}_{y_1,x_1x_3,y_2,x_2x_4}.
\end{equation}
In the following, the upper suffix is omitted
for simplicity. The two introduced tensors are first decomposed using the singular value decomposition (SVD) as \begin{equation}
  \begin{cases}
    M_1=U_1L_1{V_1}^{\dagger}=U_1R_1, \\
    M_2=U_2L_2{V_2}^{\dagger}=U_2R_2.
  \end{cases}
  \label{eqn:SVD_M}
\end{equation}
Using the decomposition, two terms in Eq.~(\ref{eqn:epsilon}) are
expressed as
\begin{equation}
  M_1 M_2^{\dagger} = U_1 R_1  R_2^{\dagger} U_2^{\dagger},
\end{equation}
and
\begin{equation}
  M_1 P_1 P_2 M_2^{\dagger} = U_1 R_1 P_1 P_2
   R_2^{\dagger} U_2^{\dagger},
\end{equation}
respectively.

\begin{figure}
 \centering
 \includegraphics[width=\linewidth]{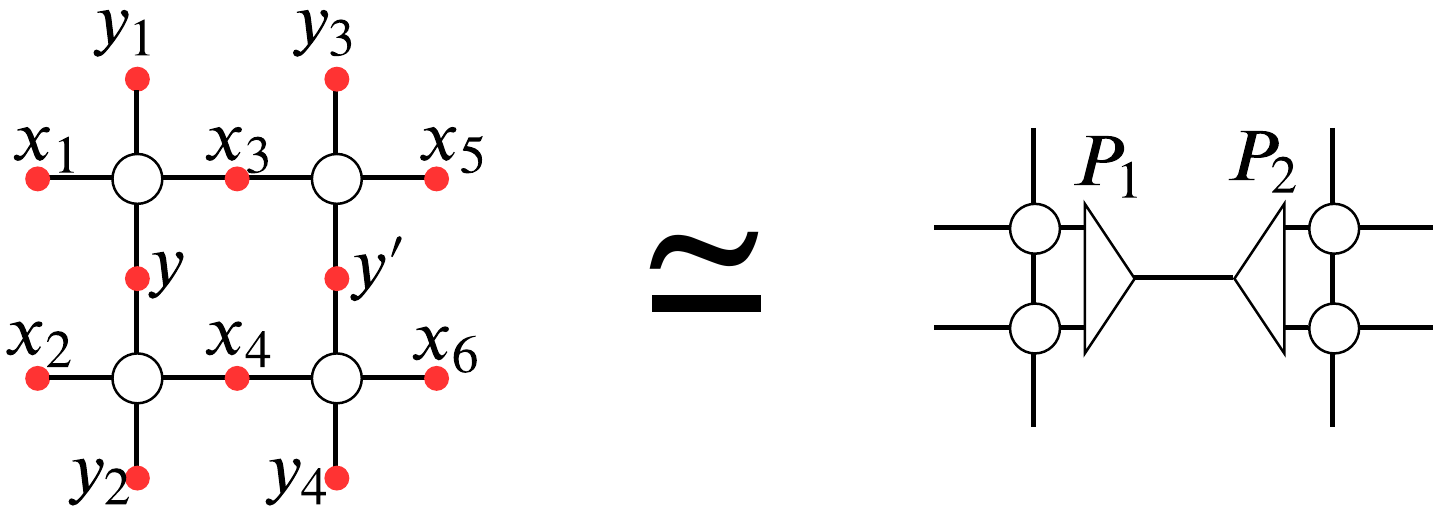}
 \caption{Two elementary diagrams in the HOTRG: a TN consisting of four tensors (left) and corresponding
 compressed network with projection tensors (right).}
 \label{fig:projector}
\end{figure}

The amount of calculation required can be reduced by performing SVD on
$M_1^{\dagger}M_1$ and $M_2^{\dagger}M_2$ as
\begin{equation}
  \begin{cases}
    M_1^{\dagger}M_1=W_1L^2_1{V_1}^{\dagger} \longrightarrow
   R_1=L_1V_1^{\dagger}, \\
    M_2^{\dagger}M_2=W_2L^2_2{V_2}^{\dagger} \longrightarrow
   R_2=L_2V_2^{\dagger},
  \end{cases}
  \label{eqn:SVD_M^2}
  \end{equation}
To compress the bond dimension between $R_1$ and $R_2^{\dagger}$ to
$D_{\text{cut}}$, SVD is again used as
\begin{equation}
  R_1 R_2^{\dagger} = U L V^{\dagger}.
  \label{eqn:SVD_R1R2}
\end{equation}
Let $L'$ be a $(D^2,D_{\text{cut}})$ array of the top $D_{\text{cut}}$ largest
singular values of $L$. The reduced tensors $R_1'$ and $R_2'$ are
defined as
\begin{equation}
  \begin{cases}
    R'_1 = U L'^{\frac{1}{2}},\\
    {R'_2}^{\dagger} = L'^{\frac{1}{2}} V^{\dagger},
  \end{cases}
  \label{eqn:R1'R2'}
\end{equation}
respectively. They provide the decomposition minimizing the error for a
given $D_{\rm cut}$.
Therefore, $P_1$ and $P_2$ which minimize Eq.~(\ref{eqn:epsilon}), satisfy the
conditions
\begin{equation}
  \begin{cases}
    R_1P_1 = U L'^{\frac{1}{2}},\\
    P_2R_2^{\dagger} = L'^{\frac{1}{2}} V^{\dagger},
  \label{eqn:projector1}
  \end{cases}
\end{equation}
Here, from Eq.~(\ref{eqn:R1'R2'}), the following equations hold:
\begin{equation}
 \begin{cases}
  R_1{R_2}^{\dagger}V L'^{-\frac{1}{2}} =  ULV^{\dagger} VL^{-\frac{1}{2}} = U
  L'^{\frac{1}{2}} =R_1P_1, \\
  L'^{-\frac{1}{2}} U^{\dagger}R_1{R_2}^{\dagger} = L'^{-\frac{1}{2}}U^{\dagger}ULV^{\dagger}
  = L'^{\frac{1}{2}}V^{\dagger} = P_2{R_2}^{\dagger}.
 \end{cases}
 \label{eqn:projector2}
\end{equation}
Eventually, the projections $P_1$ and $P_2$ that minimize the error
(\ref{eqn:epsilon}) are given by
\begin{equation}
 \begin{cases}
  P_1 = {R_2}^{\dagger} V L'^{-\frac{1}{2}}, \\
  P_2 = L'^{-\frac{1}{2}} U^{\dagger} R_1,
 \end{cases}
 \label{eqn:Projector}
\end{equation}
respectively.

\begin{figure}
 \centering
 \includegraphics[width=\linewidth]{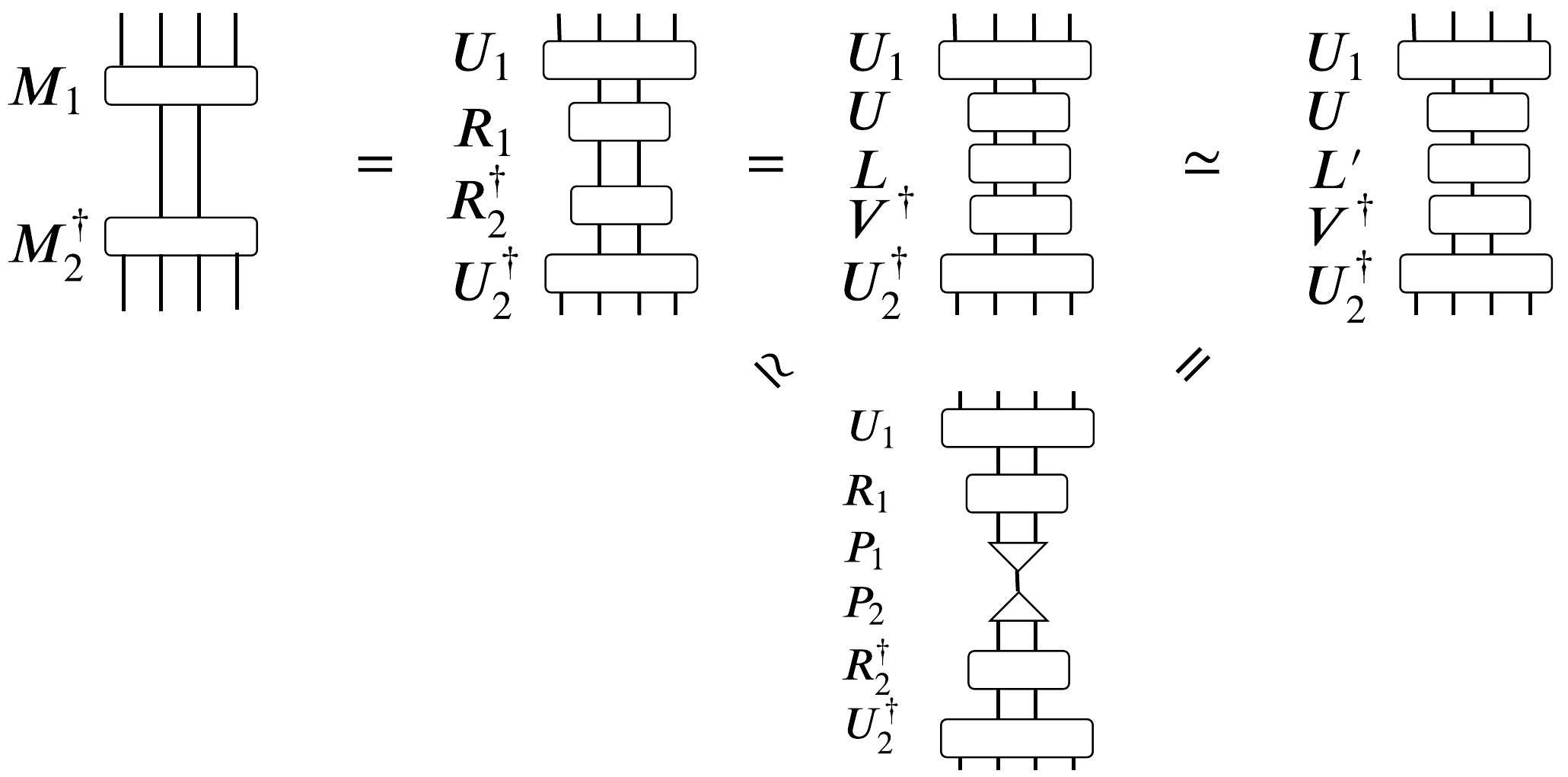}
 \caption{Diagrammatic representation of the process of construction of projection tensors $P_1$ and $P_2$ given by Eq.~(\ref{eqn:Projector}).}
 \label{fig:projector2}
\end{figure}


\end{document}